\documentclass[%
twocolumn,
 reprint,
superscriptaddress,
 amsmath,amssymb,
 aps,
 rc ]{revtex4-2}
\usepackage{comment}
\usepackage{physics}
\usepackage{graphicx}% Include figure files
\usepackage{dcolumn}% Align table columns on decimal point
\usepackage{bm}% bold math
\usepackage{xcolor}
\usepackage{siunitx}
\usepackage{chemformula}
\DeclareSIUnit\sq{\ensuremath\Box}
\DeclareSIUnit{\belmilliwatt}{Bm}
\DeclareSIUnit{\dBm}{\deci\belmilliwatt}
\usepackage{bibunits}

\usepackage{hyperref}% add hypertext capabilities
\begin{document}

\preprint{APS/123-QED}

\title{Non-Markovian dynamics of a superconducting qubit in a phononic bandgap}
%\title{Non-Markovian dynamics of a phonon-protected superconducting qubit}
%\\An atomic indistinguishable photon source in silicon photonics}% Force line breaks with \\

\author{Mutasem Odeh}
\affiliation{
	Department of Electrical Engineering and Computer Sciences, University of California,  Berkeley, Berkeley, California 94720, USA
}
\affiliation{
	Materials Sciences Division, Lawrence Berkeley National Laboratory, Berkeley, California 94720, USA
}%

\author{Kadircan Godeneli}

\affiliation{
	Department of Electrical Engineering and Computer Sciences, University of California,  Berkeley, Berkeley, California 94720, USA
}
\affiliation{
	Materials Sciences Division, Lawrence Berkeley National Laboratory, Berkeley, California 94720, USA
}%
\author{Eric Li}

\affiliation{
	Department of Electrical Engineering and Computer Sciences, University of California,  Berkeley, Berkeley, California 94720, USA
}

\author{Rohin Tangirala}

\affiliation{
	Department of Electrical Engineering and Computer Sciences, University of California,  Berkeley, Berkeley, California 94720, USA
}

\author{Haoxin Zhou}

\affiliation{
	Department of Electrical Engineering and Computer Sciences, University of California,  Berkeley, Berkeley, California 94720, USA
}

\affiliation{
	Materials Sciences Division, Lawrence Berkeley National Laboratory, Berkeley, California 94720, USA
}%
\affiliation{
Department of Physics, University of California, Berkeley, Berkeley, California 94720, USA
}

\author{Xueyue Zhang}

\affiliation{
	Department of Electrical Engineering and Computer Sciences, University of California,  Berkeley, Berkeley, California 94720, USA
}
\affiliation{
Department of Physics, University of California, Berkeley, Berkeley, California 94720, USA
}
\author{Zi-Huai Zhang}

\affiliation{
	Department of Electrical Engineering and Computer Sciences, University of California,  Berkeley, Berkeley, California 94720, USA
}
\affiliation{
	Materials Sciences Division, Lawrence Berkeley National Laboratory, Berkeley, California 94720, USA
}%
\affiliation{
Department of Physics, University of California, Berkeley, Berkeley, California 94720, USA
}

\author{Alp Sipahigil}
\email{Corresponding author: alp@berkeley.edu}

\affiliation{
Department of Electrical Engineering and Computer Sciences, University of California,  Berkeley, Berkeley, California 94720, USA
}
\affiliation{
 Materials Sciences Division, Lawrence Berkeley National Laboratory, Berkeley, California 94720, USA
}%
\affiliation{
Department of Physics, University of California, Berkeley, Berkeley, California 94720, USA
}

\date{\today}% It is always \today, today,
\begin{abstract}

% %Abstracts of Research Articles and Reports should explain to the general reader why the research was done, what was found and why the results are important. They should start with some brief BACKGROUND information: a sentence giving a broad introduction to the field comprehensible to the general reader, and then a sentence of more detailed background specific to your study. This should be followed by an explanation of the OBJECTIVES/METHODS and then the RESULTS. The final sentence should outline the main CONCLUSIONS of the study, in terms that will be comprehensible to all our readers. The Abstract is distinct from the main body of the text, and thus should not be the only source of background information critical to understanding the manuscript. Please do not include citations or abbreviations in the Abstract. The abstract should be 125 words or less. For Perspectives and Policy Forums please include a one-sentence abstract.
% \alp{a sentence giving a broad introduction to the field comprehensible to the general reader, and then a sentence of more detailed background specific to your study. This should be followed by an explanation of the OBJECTIVES/METHODS and then the RESULTS. The final sentence should outline the main CONCLUSIONS of the study, in terms that will be comprehensible to all our readers.}
%

The overhead to construct a logical qubit from physical qubits rapidly increases with the decoherence rate. Current superconducting qubits reduce dissipation due to two-level systems (TLSs) by using large device footprints. However, this approach provides partial protection, and results in a trade-off between qubit footprint and dissipation. This work introduces a new platform using phononics to engineer superconducting qubit-TLS interactions. We realize a superconducting qubit on a phononic bandgap metamaterial that suppresses TLS-mediated phonon emission. We use the qubit to probe its thermalization dynamics with the phonon-engineered TLS bath. Inside the phononic bandgap, we observe the emergence of non-Markovian qubit dynamics due to the Purcell-engineered TLS lifetime of \SI{34}{\micro \second}. We discuss the implications of these observations for extending qubit relaxation times through simultaneous phonon protection and miniaturization.

\end{abstract}

\maketitle
A superconducting quantum processor with practical utility requires a large number of highly coherent, error-corrected qubits \cite{EC1, EC2, EC3} to achieve quantum advantage \cite{PhysRevA.95.032338, Shore, Grover}. Scaling recent logical qubit demonstrations \cite{EC5} will require further improvements in gate and readout errors, as well as qubit footprint miniaturization.  Miniaturizing superconducting qubits while improving their coherence is a challenging task, as miniaturization often leads to increased dissipation due to stronger coupling to two-level systems (TLSs). TLSs are surface or bulk defects within disordered or amorphous solids. Their strong electric and elastic dipole moments make them the dominant dissipation channel of current superconducting qubits \cite{TLSreview, TLS2, TLS3}.  Large planar qubits can reduce TLS-induced dissipation by minimizing energy participation of lossy interfaces \cite{PhysRevLett.107.240501,10.1063/1.3637047}. While this approach enabled improvements in qubit coherence up to a few hundred microseconds, simultaneous improvements to lifetime and footprint remain as outstanding challenges for scaling \cite{irfanReview, T1qubit1, T1qubit2, T1qubit3}.

In this work, we demonstrate an alternative approach to address the qubit footprint-dissipation trade-off by using a phonon-engineered qubit with a modified TLS bath.  The electric and elastic dipole moments of TLSs mediate coupling between the superconducting qubit and the phonon bath, resulting in a phononic Purcell decay channel for the qubit. We use a phononic bandgap metamaterial as a mechanical Purcell filter to suppress the spontaneous phonon emission of TLSs, which in turn, influences the qubit relaxation time. We use the qubit to populate the TLS bath and to characterize the modified dissipative dynamics of the qubit and the TLSs \cite{TLSpulses1, TLSpulses2}. We observe a strong enhancement of TLS lifetime in the phononic bandgap, along with a signature of qubit lifetime improvement. The observations are well-modeled with Solomon equations for a qubit coupled to a TLS environment \cite{Solomon, spiecker2023solomon}, and show non-Markovian qubit dynamics inside the phononic bandgap \cite{nonMarkov1, nonMarkov2}. We discuss prospects for combining phonon protection and miniaturization to enable next-generation quantum processors.
%####################
%####################
%####################
%####################
%####################d

\begin{figure*}[t!]
	\centering
	\includegraphics[width=2\columnwidth]{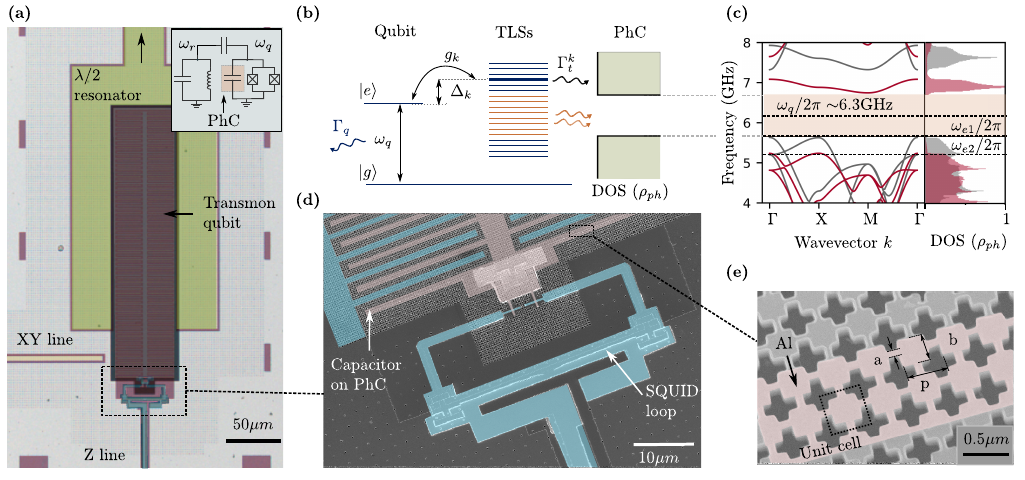}
	\caption{\textbf{A superconducting qubit on a phononic bandgap metamaterial.}
		(a) Optical micrograph (false-colored) of the transmon qubit 
  %, formed by an interdigitated capacitor and a pair of Josephson junctions (JJ) 
  fabricated on a suspended silicon phononic metamaterial. %The qubit is coupled to a readout resonator (green), an XY line for control (yellow), and a Z-line for frequency tuning (blue). 
  Inset: equivalent circuit. 
		(b)  The qubit and the $k$th TLS interact with coupling strength $g_k$ and detuning $\Delta_k$, and decay to their respective environments at rates $\Gamma_q$ and $\Gamma_t^k$.
		(c) Simulated band structure and density of states (DOS) of the fabricated phononic crystal (PhC). Lower band edges $\omega_{e1}/2\pi=$~\SI{5.6}{\giga\hertz} and $\omega_{e2}/2\pi$~=~\SI{5.2}{\giga\hertz}  are for the \SI{190}{\nano\meter}-thick Si (gray) and $220/50$\SI{}{\nano\meter}-thick Si/Al (red) unit cells. %The energy bands of \SI{190}{\nano\meter}-thick Si unit cell has a bottom band edge $\omega_{e1}/2\pi=$ \SI{5.6}{\giga\hertz} (gray) where the bands for $220/50$\SI{}{\nano\meter}-thick Si/Al unit cell has a bottom edge $\omega_{e2}/2\pi$ = \SI{5.2}{\giga\hertz} (red). 
        The complete phononic bandgap is shaded in brown. Maximum transmon frequency $\omega_q/2\pi \approx $ \SI{6.3}{\giga\hertz}.
		(d,e) False-colored scanning electron micrographs of the SQUID loop and interdigitated capacitor on the phononic crystal. The device tested in this work does not have a phononic crystal around the Josephson junctions.
		(e) Unit cell dimensions: $\{a,b,p\}=\,$ \{70, 320, 445\}\,\SI{ }{\nano \meter}.}
		
	\label{fig1}
\end{figure*}
\noindent{}\textbf{A superconducting qubit on a phononic bandgap metamaterial.} According to the standard tunneling model, TLSs display strong electric ($\sim$1 Debye) and elastic ($\sim$\SI{1}{\electronvolt}) dipole transition matrix elements that result in their strong interactions with superconducting circuits and phonons \cite{TLSreview,ph1,ph2}. Their linear electromechanical response can be considered as atomic-scale piezoelectricity that converts the energy of superconducting qubits to phonons. As illustrated in Fig.~\ref{fig1}(b), a TLS couples to the oscillating electric field of the qubit and dissipates it to the substrate via phonon emission at a rate $\Gamma_t^k$. Consequently, the qubit experiences Purcell decay through each TLS with a rate $\Gamma_{qt}^k$. When higher order coherent effects can be ignored, the total qubit decay rate is then the sum of its intrinsic decay rate $\Gamma_q$  (from non-TLS sources) and the Purcell decay rates due to  the TLS ensemble:
\begin{equation}
	\Gamma_{1} = \Gamma_q + \sum_{k}\frac{2g_k^2 \Gamma_{m}}{\Gamma_{m}^2+\Delta_k^2}
	\label{crossRelaxtionEq}
\end{equation}
where $\Delta_k$ is the detuning between the qubit and the $k$th TLS, $g_k$ is their transverse coupling strength, and $\Gamma_{m} = (\Gamma_q + \Gamma_t^k)/2$ is the mutual decoherence rate in the absence of dephasing \cite{Solomon}.

%for $\Gamma_m^2 + \Delta_k^2 \gg \abs{g_k}^2$. 

The TLS lifetime is determined by the spontaneous phonon emission ($\Gamma_t^k$) rate and can be improved by suppressing the phonon density of states. If the spectral density of the phonon-protected TLS ($\rho$) is much smaller than mutual decoherence time ($1/\Gamma_m$), the decay rates of the qubit ($\Gamma_1$) and TLSs ($\Gamma_t^k$) are closely linked \cite{Rosen}. In this regime, the qubit decay $\Gamma_1$ is also expected to be suppressed, and the phononic crystal can be viewed as a mechanical Purcell filter that suppresses TLS-induced phonon emission from a superconducting qubit.

We study the modified qubit-TLS interactions on a phononic metamaterial with an engineered phonon density of states. Our device consists of an all-aluminum tunable transmon qubit on a suspended 2D phononic crystal membrane (Fig.~\ref{fig1}(a)) \cite{SOI}. The transmon consists of a compact interdigitated capacitor shunted to the ground through two symmetric Josephson junctions that form a superconducting quantum interference device (SQUID), as shown in Fig.~\ref{fig1}(d). The SQUID loop is inductively coupled to a Z-control line for qubit frequency tuning. The transmon is capacitively coupled to an XY line for qubit control and to a $\lambda/2$ coplanar waveguide resonator for fast dispersive readout. The device is mounted at the mixing chamber stage ($\sim$\SI{10}{\milli \kelvin}) of a dry dilution refrigerator and enclosed in multiple layers of radiation and magnetic shields. Details about the fabrication process, experimental setup, and measured qubit parameters are provided in the supplementary materials \cite{SI}.

The qubit capacitor is formed by interdigitating \SI{1}{\micro\meter}-wide fingers with \SI{1}{\micro\meter} gaps. The capacitor is fully engraved by the underlying phononic crystal structure as shown in Fig.~\ref{fig1}(e). We design the \SI{260}{\micro\meter} $\times$ \SI{60}{\micro\meter} capacitor using an effective medium description for the dielectric constant of the phononic crystal \cite{SI}. The mass loading due to aluminum electrodes alters the phononic band structure and shifts the lower band edge from \SI{5.6}{\giga\hertz} to \SI{5.2}{\giga\hertz} (Fig.~\ref{fig1}(c)), a signature that will be visible in subsequent qubit measurements. The common bandgap is centered at \SI{6.2}{\giga\hertz} with a \SI{1.2}{\giga\hertz} bandwidth. At $\omega_q/2\pi=$ \SI{6.3}{\giga \hertz}, we measure a qubit lifetime of $T_1=$ \SI{0.42}{\micro\second} and Ramsey dephasing time of $T_2^*=$ \SI{0.61}{\micro\second}. We tune the qubit frequency from \SI{6.3}{\giga \hertz} down to \SI{4}{\giga \hertz} and do not observe any avoided level crossings with TLSs \cite{SI}. The frequency response implies that the qubit incoherently interacts with a high-density bath of weakly coupled  TLSs, consistent with the large mode volume and surface participation ratio of the design \cite{XMON}. The designed and measured qubit properties can be found in the supplementary materials \cite{SI}.

\begin{figure}[!t]
	\centering
	\includegraphics[width=\columnwidth]{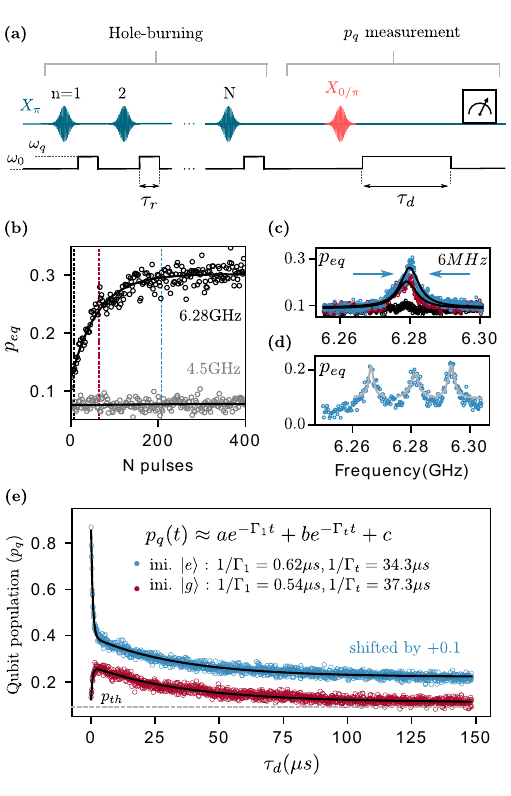}
	\caption{
	\textbf{Saturating phonon-gapped two-level systems with a qubit.} 
(a)	Sequence for hole-burning the TLS bath and measuring its dynamics with a qubit. The qubit is prepared in the excited state at $\omega_0/2\pi = \SI{6.3}{\giga\hertz}$ and is allowed to decay at $\omega_q/2\pi=\SI{6.28}{\giga\hertz}$ by waiting for $\tau_r = \SI{1}{\micro\second} > 1/\Gamma_1$. After $N$ repetitions, the thermalization dynamics of the qubit and the saturated TLS bath are measured with the qubit initialized in state $|g\rangle$ or $|e\rangle$. The TLS equilibrium population $p_{eq} \approx  p_q(\tau_d=\SI{5}{\micro \second})$ as a (b)  function of pulse number $N$ inside (black) and outside (gray) the phononic bandgap, and (c) as a function of frequency around $\SI{6.28}{\giga\hertz}$ for $N=0,50,200 $ (black, red, blue). (d) Hole-burning at three adjacent frequencies using interleaved polarization pulses.
(e) Qubit relaxation dynamics from states $|g\rangle$ (red) and $|e\rangle$  (blue) following $N=200$ polarization pulses at $ \SI{6.28}{\giga\hertz}$. Fast  ($\Gamma_1^{-1}$) and slow ($\Gamma_t^{-1}$) decay constants correspond to the qubit and TLS lifetimes. The thermal population is $p_{th} \approx 0.028$ \cite{SI}.}
	\label{fig2}
\end{figure}

%####################
%####################
%####################
%####################
%####################
\noindent{}\textbf{Qubit-driven TLS hole-burning inside a phononic bandgap.} Qubit-TLS interactions can result in coherent or incoherent dynamics depending on the ratio of interaction strength ($g_k$) to the mutual decoherence rate ($\Gamma_m$). For $|g_k|>\Gamma_m$, coherent qubit-TLS oscillations can be observed using swap spectroscopy with individual TLSs~\cite{coherentExchange}. For $|g_k|<\Gamma_m$ or in the case of dense TLS bath (approaching the continuum limit), the qubit and TLS population dynamics follow the Solomon rate equations \cite{spiecker2023solomon}:

\begin{align}
	\dot{p}_q&=-\Gamma_q(p_q-p_{th})-\sum_{k} \Gamma_{qt}^k (p_q-p_t^k)\label{SolomonEq1}\\
	\dot{p}_t^k&=-\Gamma_t(p_t^k-p_{th})- \Gamma_{qt}^k (p_t^k-p_q) 
	\label{SolomonEq2}
\end{align}
where $p_q, p_t, p_{th}$ refer to the qubit, TLS, and thermal excited state populations. In this regime, the average TLS lifetime can be measured by using a qubit-driven hole-burning sequence where we first excite the TLS bath using the qubit, and infer the bath properties from the qubit-bath thermalization dynamics. The hole-burning sequence in Fig.~\ref{fig2}(a)  consists of $X_\pi$ pulses that prepare the qubit in the $|e\rangle$ state at a reference frequency $\omega_0$. The excitation is subsequently exchanged with the TLS environment at $\omega_q$ by letting the qubit relax ($\tau_r > 1/\Gamma_1$). After $N$ hole-burning pulses, we use the qubit to probe the qubit-bath thermalization dynamics as they resonantly interact at frequency $\omega_q$ for duration $\tau_d$.

 \begin{figure*}[!t]
	\centering
	\includegraphics[width=2\columnwidth]{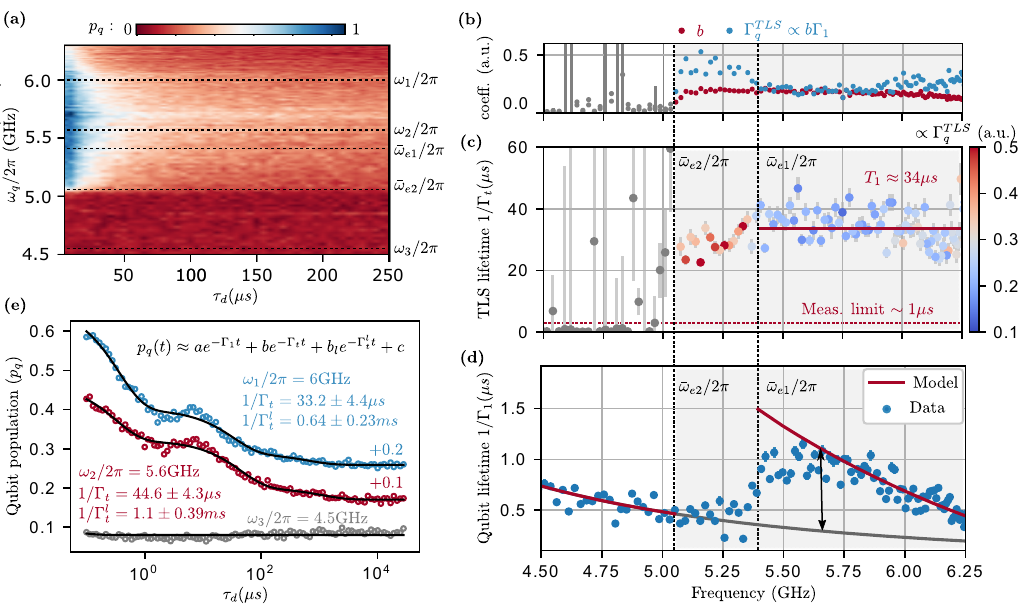}    
	\caption{\textbf{Non-Markovian dynamics of a phonon-protected qubit.}
(a)	Qubit relaxation dynamics  inside ($\omega>\bar{\omega}_{e2}$) and outside ($\omega<\bar{\omega}_{e2}$) the phononic bandgap after $N = 200$ polarization pulses. Experimentally extracted band edges: $\bar{\omega}_{e2}/2\pi=$\SI{5.4}{\giga\hertz} and $\bar{\omega}_{e1}/2\pi=$ \SI{5.05}{\giga\hertz}.  
Linecuts of the data are fit to the model in Fig.~\ref{fig2}(e) to infer: (b) amplitude $b$ (red) and TLS-induced qubit decay rate $\Gamma_q^{TLS} \propto b \Gamma_1$ (blue), (c)  	TLS lifetime ($1/\Gamma_t$), and (d) qubit lifetime ($1/\Gamma_1$). 
%The amplitude of the TLS decaying exponent $b$ (red) and the qubit decay due to TLSs $\Gamma_q^{TLS} \propto b \Gamma_1$ (blue). 
%The increase in $\Gamma_q^{TLS}$ when moving from $\bar{\omega}_{e1}$ to $\bar{\omega}_{e2}$ indicates that a subset of TLSs become short-lived as they leave their local bandgap.
(c)	The average TLS lifetime is \SI{34}{\micro\second} inside the phononic bandgap ($\omega>\bar{\omega}_{e1}$), and drops below the detection limit ($<$\SI{1}{\micro\second}) outside the bandgap ($\omega<\bar{\omega}_{e2}$). Color-coding represents $\Gamma_q^{TLS}$. 
(d) Qubit lifetime undergoes a smooth increase at the band edge $\bar{\omega}_{e1}$. The model (red, \cite{SI}) suggests around twofold qubit lifetime improvement when compared with the model prediction without a phononic bandgap (gray).
(e) Slow-relaxation dynamics of the qubit for $\{\omega_1,\omega_2,\omega_3\}/2\pi=\{6, 5.6, 4.5\}$ \SI{}{\giga \hertz} measured up to \SI{20}{\milli\second}. A tri-exponential model captures the additional slow decay timescale $\Gamma_{t}^l$ for the TLS bath. 
}
\label{fig3}
\end{figure*}
For $\Gamma_1 \gg \Gamma_t$, the qubit-TLS bath thermalization rate is faster than that of the TLSs to the phonon bath. In this regime, the qubit population approximates the equilibrium TLS population for $\tau_d \gg 1/\Gamma_{1}$. Fig.~\ref{fig2}(b) shows the TLS population as a function of the number of hole-burning pulses  $N$. Around $N=200$, the TLS bath can be populated from its thermal state to around $p_{eq}\approx30\%$ near the center of the phononic bandgap ($\sim$\SI{6.28}{\giga\hertz}, black). This response shows the presence of long-lived TLSs and the emergence of a non-Markovian bath inside the phononic bandgap. This is in stark contrast to measurements performed at frequencies outside the phononic bandgap ($\sim$\SI{4.5}{\giga\hertz}, gray data), where the TLSs cannot be populated and measured using the pulse sequence due to their very short lifetime (\SIrange{10}{100}{\nano \second} in Refs.\,\cite{PhysRevLett.105.177001, PhysRevLett.111.080502}). 

These observations can be explained by considering the steady-state populations that are determined by the TLS excitation rate through the qubit ($(1-p_t)/\tau_r$) and the energy decay rate from $\text{N}$ TLSs resonantly interacting with the qubit ($N p_t \Gamma_t$). The competition between these rates results in a steady state population $p_t \approx \Gamma_r/(\text{N}\Gamma_t + \Gamma_r)$. We use this relation to infer the effective number of TLSs resonantly interacting with the qubit  $\text{N}=100$  TLSs at \SI{6.28}{\giga\hertz}. Outside the phononic bandgap, $p_t \approx 0$ due to the fast, Markovian relaxation of the TLS bath. 

We probe the spectral distribution of the populated TLS bath, and observe a peak around the hole-burning frequency with a linewidth of $6.8\pm 0.2 $\SI{}{\mega\hertz}, which includes the dephasing of the probe qubit and the probed TLSs (Fig.~\ref{fig2}(c)). This indicates that the populated TLSs are dense ($\rho\approx~$\SI{15}{\per \mega \hertz}) and share similar Purcell decay rates. As shown in Fig.~\ref{fig2}(d), we use interleaved polarization pulses at different frequencies to saturate the TLS bath at different spectral regions, confirming the uniform, high-density distribution of the TLS bath (Fig.~\ref{fig2}(d)). Under the dense and uniform bath approximation ($\Gamma_{qt}^k\approx \Gamma_{qt}$), the rate equations simplify to \cite{spiecker2023solomon, SI}
\begin{equation}
	\dot{p}_q=-\Gamma_1(p_q-p_{th})+\Gamma_q^{TLS} p_{t,0}^* e^{-\Gamma_t t}.
	\label{diffEqtext}
\end{equation}
where $\Gamma_q^{TLS}=\sum_{k}\Gamma_{qt}^k$, and $p_{t,0}$ represents the initial TLS population. For long-lived TLSs ($\Gamma_t\ll \Gamma_{1}$), the solution of the differential equation can be approximated by a biexponential form $p_q(t)\approx a e^{-\Gamma_1 t} + b e^{-\Gamma_t t} + p_{th}$~\cite{SI}. In Fig.~\ref{fig2}(e), we probe the qubit decay after $N=200$ polarization pulses with the qubit initialized in the $|g\rangle$ and $|e\rangle$ states. By fitting the data to a biexponential form, we observe that, on average, the qubit lifetime is $1/\Gamma_1=1/(\Gamma_{\uparrow}+\Gamma_{\downarrow})\approx$ \SI{0.58}{\micro\second}, where $\Gamma_{\uparrow} (\Gamma_{\downarrow}$) is the upward (downward) transition rate and depends on the TLS population. However, the qubit lifetime ($1/\Gamma_1$) remains independent of the TLS population, and the TLS lifetime  ($1/\Gamma_t$) is independent of the qubit initialization \cite{SI}.

%####################
%####################
%####################
%####################
%####################
\noindent{}\textbf{Non-Markovian dynamics of a phonon-protected superconducting qubit.} To probe the effectiveness of the phononic bandgap, we performed the hole-burning pulse sequence used in Fig.~\ref{fig2}(e) over the frequency range of \SIrange{4}{6.25}{\giga\hertz}, with a time delay of up to $\tau_d=$ \SI{250}{\micro\second}. This measurement captures both the slow and fast qubit dynamics from which the qubit ($1/\Gamma_1$) and TLS ($1/\Gamma_t$) lifetimes can be extracted. As shown in Fig.~\ref{fig3}(a), the qubit population at long delays rapidly vanishes when we move the qubit outside the phononic bandgap ($\omega < \bar{\omega}_{e2}$). The qubit decay due to TLSs can be estimated by noting that $\Gamma_q^{TLS} \propto b\Gamma_1$ where $\Gamma_q^{TLS}=\sum_{k}\Gamma_{qt}^k$ (Eq.\ref{eqn:A8},~\cite{SI}). We use the results presented in Fig.~\ref{fig3}(b) to experimentally locate the band edges corresponding to the Si ($\bar{\omega}_{e1}/2\pi$ = \SI{5.4}{\giga\hertz}) and Si/Al ($\bar{\omega}_{e2}/2\pi$ = \SI{5.05}{\giga\hertz}) unit cells. For $\bar{\omega}_{e2}<\omega<\bar{\omega}_{e1}$, a subset of the TLS bath leaves the bandgap and becomes short-lived, reducing both the qubit and average TLS lifetime (Fig.~\ref{fig3}(c) and (d)). The phononic bandgap improves the TLS lifetime from values below the detection limit ($<$ \SI{1}{\micro\second}) to an average of \SI{34}{\micro\second}. The qubit lifetime (Fig.~\ref{fig3}(d)) shows frequency dependence and experiences a smooth increase at the band edge due to the suppression of the TLS decay. An approximate model of the qubit-TLS interaction can be constructed from the TLS lifetimes and Eq.~\ref{crossRelaxtionEq}, from which we estimate a coupling strength $g/2\pi\sim$ \SI{50}{\kilo \hertz}, TLS density $\rho \sim$ \SI{20}{\per \mega \hertz} (in good agreement with the estimation from Fig.~\ref{fig2}(c)), and the intrinsic qubit decay rate $\Gamma_q/2\pi\sim$ \SI{5}{\kilo \hertz} \cite{SI}.

We observe a small residual population inside the bandgap that decays far beyond $\tau_d=$ \SI{250}{\micro\second} suggesting the presence of a distribution of lifetimes for the TLS bath. To probe the long lifetime, we measured the relaxation dynamics of the qubit up to \SI{20}{\milli\second} for three different frequencies as shown in Fig.~\ref{fig3}(e). By fitting the data to a tri-exponential form $p_q(t) \approx a e^{-\Gamma_1 t} + b e^{-\Gamma_{t} t} + b_l e^{-\Gamma_{t}^l t} + c$, we extract two major time scales of the TLS bath. We measured $1/\Gamma_{t} (1/\Gamma_{t}^l$) of \SI{33.2}{\micro \second} (\SI{0.64}{\milli \second}) for $\omega_1/2\pi=$ \SI{6}{\giga \hertz} and of \SI{44.6}{\micro \second} (\SI{1.1}{\milli \second}) for $\omega_2/2\pi=$ \SI{5.6}{\giga \hertz}. Further investigation (detailed in \cite{SI}) indicates that $1/\Gamma_{t}^l$ can increase further to \SI{1.67}{\milli \second} (\SI{2.8}{\milli \second}) for $\omega_1 (\omega_2)$ if the qubit is detuned from the TLS population during the time delay $\tau_d$, suggesting that the qubit Purcell limits the TLS lifetime at such long timescales. 

%####################
%####################
%####################
%####################
%####################
\noindent{}\textbf{Discussion and outlook.}
We showed that embedding a superconducting qubit inside a phononic bandgap enhances the TLS bath lifetime and results in non-Markovian qubit dynamics. The TLSs inside the phononic bandgap exhibited relaxation times ranging from \SI{34}{\micro \second} to \SI{1.1}{\milli \second}, extending up to \SI{2.8}{\milli \second} when the qubit is detuned. Outside the bandgap, the relaxation time is less than \SI{1}{\micro \second} (limit of detection), consistent with the \SIrange{10}{100}{\nano \second} measured values reported in \cite{PhysRevLett.105.177001, PhysRevLett.111.080502}. While the precise mechanism limiting the TLS lifetime is currently unknown and subject to further study, we expect that lower disorder phononic metamaterials with larger bandgaps would result in further improvements to TLS lifetimes and suppress non-resonant relaxation mechanisms \cite{moe,msphononRes}.

%The TLSs inside the phononic bandgap showed a relaxation time ranging from \SI{34}{\micro \second} to \SI{1.1}{\milli \second} and up to \SI{2.8}{\milli \second} when the qubit is detuned. Outside the bandgap, the relaxation time is less than \SI{1}{\micro \second} (limit of detection), in good agreement with the \SIrange{10}{100}{\nano \second} measured values reported in \cite{PhysRevLett.105.177001, PhysRevLett.111.080502}. 

\begin{figure}[!t]
	\centering
	\includegraphics[width=\columnwidth]{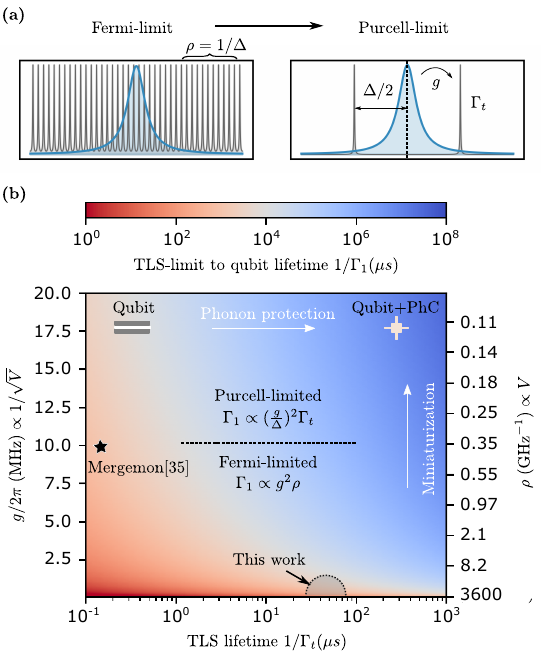}    
	\caption{\textbf{Qubit-TLS interaction regimes.}
	(a) Illustrative model of qubit (blue) and TLS (gray) spectra in the Fermi and Purcell limits. (b) TLS-limited qubit relaxation time $1/\Gamma_1$ calculated using the analytical sum of Eq.~\ref{crossRelaxtionEq} and assuming $\Gamma_q=0$, $g \propto 1/\sqrt{V}$ and $\rho \propto V$, where $V$ is the dielectric volume. The parameters assume full participation ratio \ch{Al/AlO_x/Al} Mergemon qubit from Ref.~\cite{Mergmon} as detailed in the supplementary material \cite{SI}. Reducing the qubit size transitions the relaxation rate from being Fermi-limited (independent of $\Gamma_t$) to being Purcell-limited (proportional to $\Gamma_t$). The model predicts significant improvements in the TLS-limited qubit lifetime by combining miniaturization and phonon shielding (upper right-hand corner).} 
	\label{fig4}
\end{figure}

An intriguing question is whether our phonon engineering approach, which leads to enhancements in TLS lifetimes, can also improve the lifetimes of future superconducting qubits. The qubit relaxation rate and its relation to TLS lifetimes are governed by Eq.~\ref{crossRelaxtionEq}, exhibiting two main limits that are illustrated in Fig.~\ref{fig4}(a). The \textit{Fermi-limit} is reached when the density of TLSs is large and approaches the continuum. In this regime, the TLS relaxation rate does not have a strong influence on the qubit lifetime, and the qubit decay rate follows that of Fermi's golden rule, $\Gamma_q^{TLS} \propto g^2\rho$. The \textit{Purcell-limit} is reached when the density of TLSs is small, and the relevant TLS distribution can be truncated to a few nearest neighbors. In this limit, the TLS relaxation rate has a direct influence on the qubit lifetime, and the qubit decay rate follows the Purcell formula, $\Gamma_q^{TLS} \propto (g/\Delta)^2\Gamma_t$.

Our experiment is in the Fermi-limit where we observe only a modest improvement in the qubit lifetime inside the bandgap (Fig.~\ref{fig3}(d)) due to the high TLSs density ($\rho \approx$ \SI{20}{\per \mega \hertz}). Commonly used qubits also operate in the Fermi-limit regime and mitigate TLS loss through optimized fabrication processes (reduce $\rho$) and large planar geometries that reduce TLS coupling strength (reduce $g$) and energy participation ratio \cite{XMON, floating}. Therefore, phononic shielding is not expected to significantly improve their lifetime. However, phonon-shielding is expected to be more effective for miniaturized qubits operating in the \textit{Purcell-limit} with low TLS density. We take the recently developed \ch{Al/AlO_x/Al} merged element transmon (Mergemon) as an example \cite{Mergmon}, which has near unity energy participation ratio in the thin \ch{AlO_x} dielectric layer, and a reported TLS density per unit volume of $\rho_0 = \SI{100}{\per\cubic\micro\meter\per\giga\hertz}$ \cite{XMON}. This allows us to establish a relation between the TLS density and coupling strength ($\rho \approx 1.425\times 10^6[\text{Hz}]/g^2$), for varying Mergemon dimensions while keeping the qubit capacitance fixed. We also assume a uniform distribution of the TLSs, which allows for an analytical expression of the sum in Eq.~\ref{crossRelaxtionEq}~\cite{SI}. In Fig.~\ref{fig4}(b), the Mergemon relaxation time ($1/\Gamma_1$) is plotted as a function of the TLS relaxation time ($1/\Gamma_t$) and coupling strength ($g$), assuming no intrinsic loss ($\Gamma_q=0$). The color map clearly illustrates the Fermi-limited and Purcell-limited regimes, showing quantitatively the influence of the TLS lifetime on the qubit lifetime in each regime
Our numerical calculations validate the limited improvement of the qubit lifetime in this work ($g/2\pi \approx$ \SI{0.05}{\mega \hertz}, $\rho \approx$ \SI{20}{\per \mega \hertz}), and also affirm the $\sim$\SI{100}{\micro \second} reported Mergemon lifetime with ($g/2\pi \approx$ \SI{10}{\mega \hertz}, $\rho \approx$ \SI{0.4}{\per \giga \hertz}) \cite{Mergmon}. The Mergemon with \SI{1.4}{\micro \meter} area clearly operates in the Purcell limit, and phonon shielding is expected to improve its TLS-limited lifetime to the millisecond timescale. 
%In addition, it can allow the operation of the qubit at higher frequencies, which is otherwise typically avoided due to the enhanced coupling to TLSs ($g\propto \sqrt{\omega_q}$).
We note that the impact of the phononic crystal on the TLS and qubit Ramsey coherence time $T_2^*$ is currently unknown. %The dephasing of high-frequency TLSs are dominated by the longitudinal coupling with many low-frequency TLS fluctuators ($\hbar \omega < k_B T$), resulting in telegraphic noise and spectral diffusion \cite{TLSreview}. 
We expect that the emergence of non-Markovian qubit-TLS dynamics and long relaxation times via phonon engineering will yield similar improvements in coherence via dynamical decoupling~\cite{DD1,DD2}.
%Nonetheless, engineered dynamical decoupling sequences can be used to overcome both decoherence and operational errors, particularly in non-Markovian environments with a long correlation time scale \cite{DD1,DD2}. 
%This indicates a promising path towards scaling down qubits while improving their relaxation time. 

%Finally, the work demonstrated that the Solomon's equations and the polarization pulse sequences are powerful tools for characterizing an ensemble of incoherent TLSs and can be used to study novel non-Markovian phenomena such as superradiance and state revivals \cite{PhysRevLett.104.143601, PhysRevA.87.052323}.

%####################
%####################
%####################
%####################
%####################

\noindent{}\textbf{Acknowledgments.} This work was primarily funded by the U.S. Department of Energy, Office of Science, Office of Basic Energy Sciences, Materials Sciences and Engineering Division under Contract No. DE-AC02-05-CH11231 in the Phonon Control for Next-Generation Superconducting Systems and Sensors FWP (KCAS23). Additional support was provided for device fabrication by the ONR and AFOSR Quantum Phononics MURI program. The devices used in this work were fabricated at UC Berkeley's NanoLab. We thank Irfan Siddiqi and his group for assistance with Josephson junction and bandaid fabrication. The open-source Python package QICK \cite{QICK} was used for measurements, QuTiP and SCQubit for device simulations \cite{QuTiP,SCQubit}, and Qiskit-Metal \cite{qiskit-metal} for GDS generation.

%\noindent{} \textbf{Author Contributions}
%https://credit.niso.org

%\onecolumngrid

\bibliography{main}% Produces the bibliography via BibTeX.
\clearpage 

\tableofcontents
\appendix

\renewcommand{\thefigure}{S\arabic{figure}}
\setcounter{figure}{0}

%####################
%####################
%####################
%####################
%###################

\section{Analytical modeling}

In this section, we review the Solomon equations, which are rate equations that link the qubit population (central spin) to the population of a discrete ensemble of two-level systems (spin environment). Next, we review the limit of the Purcell decay rate under a uniform distribution of TLSs and constant coupling, illustrating the Purcell-limited and Fermi-limited regimes described in the text. We use the model to fit the qubit lifetime measurement presented in Fig.~\ref{fig3}(d), as well as to compute Fig.~\ref{fig4} and identify the Fermi-limit and Purcell-limit regimes. The detailed analysis can be found in Refs.~\cite{Solomon, spiecker2023solomon}.

\subsection{The Solomon rate equations}
\label{A1}
The system is modeled assuming the qubit with population $p_q$ is coupled to a countable number of TLSs with populations $p_t^k$. We denote $\Gamma_q$ and $\Gamma_t^k$ as the intrinsic relaxation rates of the qubit and the $k$th TLS, respectively. The Purcell decay rate $\Gamma_{qt}^k$ is given by
\begin{equation}
\Gamma_{qt}^k=\frac{2g_k^2 \Gamma_{m}}{\Gamma_{m}^2+\Delta_k^2},
\end{equation}
where $\Delta_k$ is the detuning between the qubit and the $k$th TLS, $g_k$ represents their transverse coupling strength, and the mutual decoherence is described by $\Gamma_{m}=(\Gamma_q+\Gamma_t^k)/2$ in the absence of dephasing. In the limit where the mutual decoherence of the qubit and TLS is sufficiently strong ($\Gamma_{m}>g_k$), the interaction is incoherent, and the population dynamics are governed by the Solomon rate equations:
\begin{align}
\dot{p}_q&=-\Gamma_q(p_q-p_{th})-\sum_{k} \Gamma_{qt}^k (p_q-p_t^k)\label{qpopulation}\\
\dot{p}_t^k&=-\Gamma_t(p_t^k-p_{th})- \Gamma_{qt}^k (p_t^k-p_q) \label{TLSpopulation}
\end{align}
From the initial conditions, one can determine the upward and downward transition rates of the qubit
\begin{equation}
\Gamma_\uparrow(t)=\dot{p}_q(t)|_{p_q=0} \text{ and } \Gamma_\downarrow(t)=-\dot{p}_q(t)|_{p_q=1},
\end{equation}
from which, the qubit decay rate and its equilibrium population can be determined
\begin{align}
\Gamma_1&=\Gamma_\uparrow(t)+\Gamma_\downarrow(t)=\Gamma_q+\sum_{k}\Gamma_{qt}^k \\
p_{eq}(t)&=\frac{\Gamma_\uparrow(t)}{\Gamma_1}=\frac{\Gamma_q p_{th}+\sum_{k}\Gamma_{qt}^k p_t^k(t)}{\Gamma_1}.
\end{align}
In the special case of identical Purcell decay rates ($\Gamma_{qt}^k=\Gamma_{qt}$) with a large number of TLSs, the qubit population exhibits biexponential decay behavior.
\begin{equation}
	\dot{p}_q=-\Gamma_1(p_q-p_{th})+\Gamma_q^{TLS} p_{t,0}^* e^{-\Gamma_t t},
	\label{diffEq}
\end{equation}
where $\Gamma_q^{TLS}=\sum_{k}\Gamma_{qt}^k$ is the sum of the Purcell decay rates, and $p_{t,0}$ represents the initial TLS population. It can be shown that in the case of long-lived TLSs ($\Gamma_t \ll \Gamma_1$), an approximate solution to the above differential equation is a biexponential with fast and slow decay parts that encode the qubit and TLS relaxation rates, respectively. The slowly varying amplitude of population decay for the TLSs can be obtained by setting $\dot{p}_q=0$ (adiabatic elimination), from which the approximate solution to Eq.~\ref{diffEq} can be obtained:
\begin{equation}
	\boxed{p_q(t)\approx p_{q,0}^* e^{-\Gamma_1 t}  +\frac{\Gamma_q^{TLS}}{\Gamma_{1}}  p_{t,0}^* e^{-\Gamma_t t}+p_{th}}
    \label{eqn:A8}
\end{equation}

\subsection{Purcell decay rate of  uniformly distributed TLSs}
\label{A2}
When the TLSs are spread in frequency and equally spaced by a period $\Delta$ with a unified coupling strength $g$ and mutual decoherence rate $\Gamma_m$, an analytical expression for the Purcell decay rate can be obtained as follows:
\begin{align}
	\Gamma_q^{TLS}&=\sum_{k}\frac{2g^2 \Gamma_m}{\Gamma_m^2+\Delta_k^2}\\
	&=\sum_{h=-\infty}^{\infty}\frac{ab^2}{b^2+(h-bc)^2}\\
	\Gamma_q^{TLS}&=\pi ab \frac{\sinh(2\pi b)}{\cosh(2\pi b)-\cos(2\pi bc)} \label{sinhsum}
\end{align}
where $a=2g^2/\Gamma_m$, $b=\Gamma_m/\Delta$, $c=\Delta_0/\Gamma_m$ with $\Delta_0$ being the shift of the periodic TLS with respect to the qubit and can take any value between $\Delta_0 \in \{0,\Delta/2\}$. In the limit of sparse TLSs ($b\rightarrow 0$),  the sum can be terminated to the few nearest interacting TLSs, and the decoherence follows the Purcell formula  $\Gamma_q^{TLS}\approx (g/\Delta)^2\Gamma_t$. In the limit of dense TLSs ($b\rightarrow \infty$), Eq.~\ref{sinhsum} is approximately equal to $\pi a b$ and $\Gamma_q^{TLS}\approx 2 \pi g^2 \rho$ which recovers the Fermi's golden rule and is independent of the TLS relaxation time.

\subsection{Qubit lifetime modeling}
\begin{figure}[!t]
	\centering
	\includegraphics[width=\columnwidth]{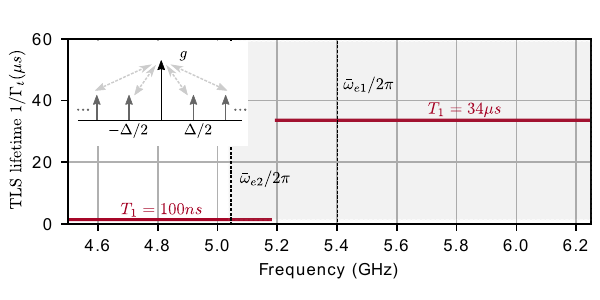}
	\caption{Qubit lifetime modeling. The TLS lifetime ($1/\Gamma_t$) is simplified to a piecewise function of \SI{34}{\micro\second} (\SI{100}{\nano\second}) inside (outside) the phononic bandgap. Inset: The model assumes that the TLSs are uniformly distributed at a constant density $\rho = 2\pi/\Delta$ with a uniform coupling strength $g_k=g$.}
	\label{fitModel}
\end{figure}

To gain more insight into the TLS bath properties, we fitted the measured qubit lifetime ($1/\Gamma_1$) of Fig.~\ref{fig3}(d) to the qubit-decay formula (Eq.~\ref{crossRelaxtionEq}). We simplified the TLS lifetime ($1/\Gamma_t$) into a piecewise function of  \SI{34}{\micro\second} for $\omega/2\pi>$ \SI{5.2}{\giga\hertz} and \SI{100}{\nano\second} otherwise, as shown in Fig.~\ref{fitModel}. The model assumes that the TLSs are uniformly distributed at a constant density $\rho = 2\pi/\Delta$ with uniform coupling strength $g$ (inset of Fig.~\ref{fitModel}). The sum of this distribution can be found analytically as in Eq.~\ref{sinhsum}.

The intrinsic qubit decay $\Gamma_q=(\frac{g_r}{\omega_r-\omega})^2 \kappa_r$ due to Purcell decay via the readout resonator is set by $\omega_r/2\pi=$ \SI{7.1}{\giga \hertz}, a decay rate of $\kappa_r/2\pi=$ \SI{2}{\mega \hertz}, and a qubit coupling strength of $g_r/2\pi=$ \SI{48}{\mega \hertz}(Appendix \ref{qubitparameters}). For the model  shown in Fig.~\ref{fig3}(d), we find an average qubit-TLS coupling strength $g/2\pi=a \omega^2$ with $a\approx$\SI{5e-11}{\per \mega \hertz} and spans a range of \SIrange{0.03}{0.08}{\mega \hertz}. The values are in good agreement with the electrostatic simulation, where $g= pE/\hbar$, and $p$ was assumed to be \SI{0.2}{e\AA} (Fig.~\ref{ESSim}(b)). However, the frequency dependence is quadratic $g\propto \omega^2$ instead of the $g\propto \sqrt{\omega}$ frequency dependence expected from the single photon amplitude. One possible reason for the discrepancy is the oversimplification of the model, where, in practice, $g$ has a complicated spatial and frequency dependence. Finally, the density of TLS per unity frequency $\rho\approx$ \SI{20}{\per \mega \hertz}, and is constant as expected from the standard tunneling model and our estimate from hole-burning sequences \cite{TLSreview}. The model fails to capture the region of the band edge, particularly between $\omega_{e1}$ and $\omega_{e2}$ (see Fig.~\ref{fig3}(d)), where a large subset of TLSs are outside of the complete bandgap.

\subsection{Case study: Merged-Element Transmon}

Here we derive the scaling between TLS density $\rho$ and coupling strength $g$ used in Fig.~\ref{fig4}(b) of \ch{Al/AlO_x/Al} using Merged-Element Transmon (Mergemon) \cite{Mergmon} as a reference. The Mergemon is modeled as a Josephson junction shunted with a parallel plate capacitor $C=\epsilon A/d$. The participation ratio is unity in the \ch{AlO_x} dielectric with electric field $E=V_{zpf}/d$, from which the coupling strength to defect TLSs with dipole moment $p$ is $g=pE/\hbar$. The TLS density per unit frequency can be expressed as:
\begin{equation}
\begin{aligned}
    \rho &=\rho_0 Ad\\
    &=\frac{\rho_0 A V_{zpf} p}{\hbar} \times \frac{1}{g}\\
    &=\frac{\rho_0 C V_{zpf}^2 p^2}{\epsilon \hbar^2 } \times \frac{1}{g^2}\\
    &\approx 1.425\times 10^6 [Hz]  \frac{1}{g^2}\\
\end{aligned}
\end{equation}
where we assumed a \SI{3.8}{\giga\hertz} qubit frequency and a capacitance $C=$ \SI{70}{\femto\farad}, from which we obtain a zero-point voltage fluctuation of $V_{zpf}=$ \SI{4}{\micro\volt}. The \ch{AlO_x} permittivity $\epsilon=10\epsilon_0$ has a TLS density per unit volume per unit frequency $\rho_0=$ \SI{100}{\per\cubic\micro\meter\per\giga\hertz} and a dipole moment of $p=$ \SI{0.2}{e\AA}. For the Mergemon design reported in \cite{Mergmon} with area $A=$ \SI{1.4}{\micro \meter \squared} and dielectric thickness $d=$ \SI{2}{\nano \meter}, we have $\rho=$ \SI{0.35}{\per \giga \hertz} and $g/2\pi\approx$ \SI{10}{\mega \hertz}.

\section{Measurements and simulations}
In this section, we provide additional measurements of the transmon qubit coherence properties and calibration. Next, we repeat the measurement presented in Fig.~\ref{fig3}(e) while detuning the qubit from the TLS bath. Lastly, we present the electrostatic simulation results from which we estimate the average qubit-TLS coupling strength ($g$).
\subsection{Qubit parameters and coherence properties}
\label{qubitparameters}

\begin{figure}[!htbp]
	\centering
	\includegraphics[width=\columnwidth]{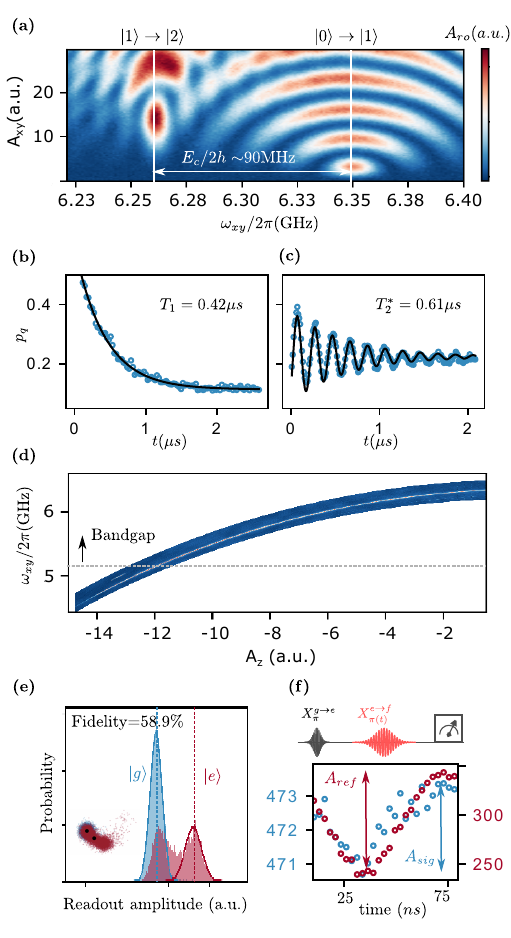}
	\caption{
	\textbf{Coherence properties of the qubit.} 
	(a) Coherent dynamics of the transmon as functions of the XY drive frequency and amplitude, from which the qubit capacitance can be estimated as $E_c/2 \approx e^2/4C_q$. (b) Qubit relaxation time $T_1 = \SI{0.42}{\micro s}$ and (c) Ramsey dephasing $T_2^* = \SI{0.61}{\micro s}$ time measured at the flux-insensitive point ($\omega_q/2\pi= \SI{6.3}{\giga\hertz}$). The relaxation time after each measurement was set to \SI{200}{\micro s} to ensure that the long-lived TLS bath relaxes to its thermal equilibrium.
	(d) Qubit two-tone spectroscopy showing a smoothly varying curve inside and outside the phononic bandgap. The absence of avoided level crossings indicates incoherent interaction with weakly coupled TLS bath.
	(e) Probability of the readout resonator amplitude for a single-shot measurement of the qubit state in the ground (blue) and excited (red) states. The readout fidelity is 58.9\% and is limited by the qubit relaxation time and the inset shows the IQ blob measurement. (f) The thermal excited-state population of the qubit $p_{th}$ measured through the $|e\rangle \rightarrow |f\rangle$ transition, as detailed in \cite{ptherm}.
}
	\label{SI:measure}
\end{figure}

We report the measured parameters as follows: $\omega_q/2\pi=$ \SI{6.3}{\giga \hertz}, $\alpha/2\pi=$ \SI{-180}{\mega \hertz}, $\omega_r/2\pi=$ \SI{7.06}{\giga \hertz}, $\Lambda/2\pi=$ \SI{4}{\mega \hertz}, and $\chi/2\pi=$ \SI{1}{\mega \hertz}. Here, $\alpha=\omega_{21}-\omega_{10}$ represents the anharmonicity which is inferred from the two-photon excitation (Fig.~\ref{SI:measure} (a)), $\omega_{r}$ is the readout resonator frequency, $2\chi=\omega_{r,\ket{0}}-\omega_{r,\ket{1}}$ denotes the dispersive shift, and $\Lambda=g^2/\Delta$ represents the Lamb shift. These measured values imply a Josephson energy $E_J/h=\SI{30}{\giga \hertz}$ in the transmon limit ($E_J\gg E_C$), where $\hbar\omega_q\approx\sqrt{8 E_J E_C}-E_C$, and a charging energy $E_C\approx -\hbar \alpha$. The readout-qubit coupling is $g_r/2\pi=\SI{55}{\mega \hertz}$, where $g\approx\sqrt{-\Delta \chi (1+\Delta/\alpha)}$, and the detuning is $\Delta=\omega_q-\omega_r$.  The readout resonator has an extrinsic quality factor of $Q_e=3.7 \times 10^3$ and an intrinsic quality factor  $Q_i=0.14-1.7 \times 10^5$ that ranges from the single photon limit to the power-saturated limit.

At $\omega_q/2\pi=$ \SI{6.3}{\giga\hertz}, we measure a relaxation time $T_1=$ \SI{0.42}{\micro\second} and Ramsey dephasing time $T_2^*=$ \SI{0.61}{\micro\second}, as shown in Fig.~\ref{SI:measure}(b) and (c), respectively. The qubit can be tuned continuously from \SIrange{6.3}{4}{\giga\hertz} with no signatures of swapping or splitting, suggesting incoherent interaction with the TLS environment (Fig.~\ref{SI:measure}(d)). We note that the Purcell decay through the readout resonator is below $1/$\SI{13}{\per \micro\second } and does not limit our coherence measurements. 

Finally, the qubit population measurements presented in the manuscript are normalized to the fitted readout amplitude distribution when the qubit is prepared in the ground and excited states, respectively, as shown in Fig.~\ref{SI:measure}(e). To identify read-out induced errors and obtain an accurate estimate of the thermal excited-state population of the qubit ($p_{th}$), we record the amplitudes $A_{ref}$ and $A_{sig}$ of the $|e\rangle \rightarrow |f\rangle$ Rabi-oscillations with and without a reference $X_{\pi}^{g\rightarrow e}$ pulse respectively, as described in \cite{ptherm}. The Rabi-oscillations are plotted in Fig.~\ref{SI:measure} (f), from which $p_{th} \approx A_{sig} / (A_{ref} + A_{sig})$, which is approximately $2.8\%$ in our device.

\subsection{Qubit-detuned TLS decay}
\begin{figure}[!t]
	\centering
	\includegraphics[width=\columnwidth]{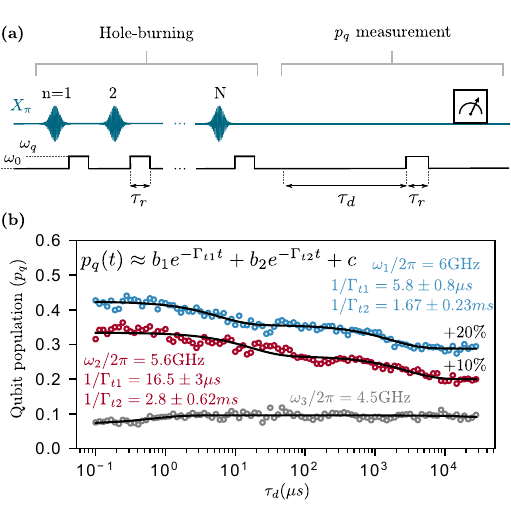}
	\caption{Qubit-detuned TLS decay: (a) Pulse sequence for TLS hole-burning and population measurement. The qubit is prepared in the excited state at $\omega_0/2\pi = \SI{6.3}{\giga\hertz}$ and is allowed to decay at $\omega_q$ by waiting for $\SI{1}{\micro\second}$. After the sequence is repeated $N$ times, the qubit is detuned to $\omega_0$ for $\tau_d$, followed by thermalization at $\omega_q$ for \SI{1}{\micro\second} and qubit readout at $\omega_0$. (b) Long-relaxation dynamics of the qubit for $\{\omega_1,\omega_2,\omega_3\}/2\pi=\{6, 5.6, 4.5\}$ \SI{}{\giga \hertz}, measured up to \SI{20}{\milli\second} and plotted on a logarithmic scale. The curve is fitted to a biexponential form, and the lifetimes are provided in the inset.}
	\label{detunedQUBTTLS}
\end{figure}

The relaxation time of the TLS ensemble can be limited by Purcell decay through the qubit. This effect becomes pronounced when measuring a small number of long-lived TLSs ($N/\Gamma_q \ll 1/\Gamma_{t}^l$). To study this effect, we used a modified TLS hole-burning sequence where the qubit is detuned from the TLS frequency $\omega_q$ throughout the duration $\tau_d$, allowing the TLS bath to decay independently, as shown in Fig.~\ref{detunedQUBTTLS} (a). At the end of the sequence, the qubit is allowed to thermalize with the TLS population for $\tau_r$, followed by a qubit readout measurement. The sequence was performed at the same frequency points of Fig.~\ref{fig3}(e), and the results were fitted to the biexponential form $p_q(t) \approx b_1 e^{-\Gamma_{t1} t} + b_2 e^{-\Gamma_{t2} t} + c$. Detuning the qubit increased the long-TLS lifetime $1/\Gamma_{t2}$ from \SI{0.64}{\milli\second} to \SI{1.67}{\milli\second} at $\omega_1/2\pi = \SI{6}{\giga\hertz}$, and from \SI{1.1}{\milli\second} to \SI{2.8}{\milli\second} at $\omega_2/2\pi =$ \SI{5.6}{\giga\hertz}. This suggests that the TLS and the qubit can Purcell-limit each other's lifetime. 

%Further studies and advanced modeling are needed to capture the TLS lifetime distribution, which becomes significant at long timescales.

\begin{figure}[!t]
	\centering
	\includegraphics[width=\columnwidth]{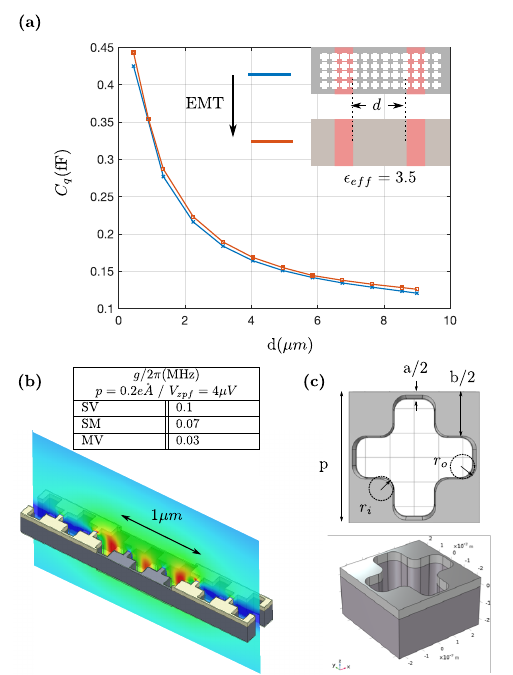}
	\caption{Electrostatic simulations: (a) Effective medium description of the phononic crystal extracted by simulating the capacitance ($C_q$) between two metal electrodes separated by $d$, for both the phononic crystal and a homogeneous slab with $\epsilon_{eff}$ = 3.5; (b) Average Qubit-TLS coupling strength ($g=pE/\hbar$), computed in \SI{3}{\nano\meter}-thick substrate-vacuum (SV), substrate-metal (SM), and metal-vacuum (MV) interfaces; and (c) Simulated phononic bandgap unit cell with parameters $a = \SI{70}{\nano \meter}$, $b = \SI{320}{\nano \meter}$, $p = \SI{445}{\nano \meter}$, $r_i = \SI{47}{\nano \meter}$, and $r_o = \SI{29}{\nano \meter}$.}
	\label{ESSim}
\end{figure}

\subsection{Electrostatic simulation }

The vast scale difference between the features of the phononic crystal (\SI{50}{\nano\meter}) and the qubit size (\SI{260}{\micro\meter} $\times$ \SI{60}{\micro\meter} capacitor) posed a problem in the electrostatic simulation of the qubit capacitance and its coupling to the readout resonator. This issue was addressed by resorting to the effective medium description of the phononic crystal. We find that an effective permittivity $\epsilon_{eff}=3.5$ of a homogeneous slab between two metal electrodes separated by spacing $d$, yields the same capacitance as that of the original phononic crystal substrate, as illustrated in Fig.~\ref{ESSim}(a). The results are in good agreement with the experimental data as well as the interdigitated capacitor simulation under periodic boundary conditions.

To estimate the average Qubit-TLS coupling strength $g$, we performed a 3D electrostatic simulation for one period of the interdigitated capacitor, as shown in Fig.~\ref{ESSim}(b). We set the voltage between the two electrodes to $V_{zpf}$=\SI{4}{\micro \volt}, which was estimated from the measured qubit parameters at $\omega/2\pi=$ \SI{6.3}{\giga \hertz} using $V_{zpf}=\omega\sqrt{\hbar Z_T/2}$, and the transmon impedance $Z_T=(\Phi_0/\pi e)\sqrt{E_C/2E_J}$. The average electric field in a \SI{3}{\nano \meter} thick layer on the substrate-vacuum (SV), substrate-metal (SM), and metal-vacuum (MV) interfaces are computed. From this data, the coupling strength to a TLS with a dipole moment of \SI{0.2}{e\angstrom} can be obtained via $g=pE/\hbar$, and the results are summarized in the table of Fig.~\ref{ESSim}(b). The simulated phononic crystal unit cell parameters are provided in Fig.~\ref{ESSim}(c).

\section{Experimental setup and methods}
This section provides the experimental details and methods for device fabrication, cryogenic setup, and the microwave electronics used in this work.

\subsection{Device fabrication}
\label{SI:fab}

\begin{figure}[!t]
	\centering
	\includegraphics[width=\columnwidth]{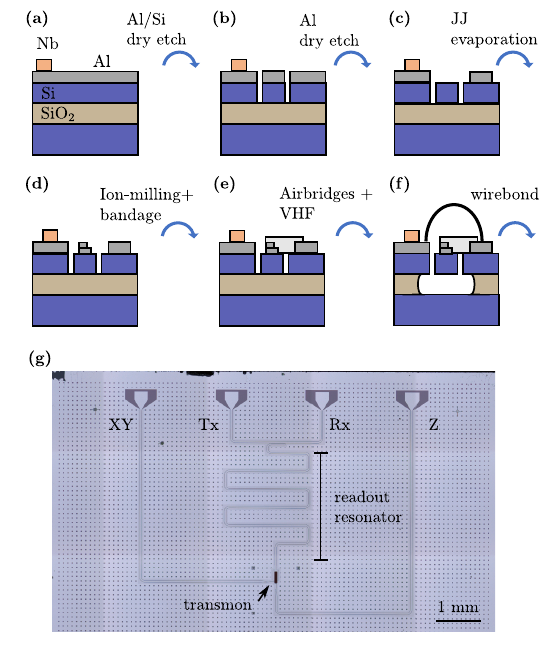}
	\caption{Schematic of the fabrication steps for the phonon-protected superconducting qubit: (a) Nb alignment marks; (b) phononic crystal electron beam lithography (EBL) and dry etching; (c) microwave circuit EBL and dry etching; (d) Josephson junction EBL and liftoff; (e) bandage EBL and liftoff; (f) airbridges followed by vapor HF device release; and (g) microscope image of the fabricated chip.}
	\label{SDevicefab}
\end{figure}

In this section, we provide the detailed fabrication process flow for the phonon-protected transmon qubit (Fig.~\ref{SDevicefab}), which is a variation of the process presented in \cite{SOI}. The SOI wafer used (supplied by Shin-Etsu) features a float zone silicon device layer with a thickness of \SI{220}{\nano\meter} and a crystal orientation of 100 ($\rho \geq$ \SI{3}{\kilo\ohm\cm}). The BOX is a \SI{3}{\micro\meter}-thick layer of \ch{SiO2} on top of a Czochralski-grown silicon handle layer with a thickness of \SI{725}{\micro\meter} ($\rho \geq$ \SI{3}{\kilo\ohm\cm}). The wafer, protected with a resist coating, is downsized from \SI{8}{''} to \SI{6}{''} (by MicroPE). Before deposition, the wafer is cleansed with \ch{H2SO4} and \ch{H2O2} (piranha solution) to remove organic residues, dipped in \ch{HCl} to remove metallic contamination, and then in \ch{HF} to remove the native oxide. Next, \SI{50}{\nm} of aluminum is sputtered at a rate of \SI{15}{\nm\per\min}. Since the contrast between materials with similar atomic numbers is poor under electron microscopy, and considering that the atomic masses of Si and Al are \SI{28}{U} and \SI{27}{U}, respectively, Nb metal (\SI{93}{U}) is used for subsequent electron-beam lithography (EBL) alignments. To define the markers, a \SI{1}{\micro \meter}-thick AZ-MIR 701 resist is exposed (Heidelberg MLA150), developed in MF-26A, and then descummed in \ch{O2} plasma. Next, a \SI{200}{\nm}-thick \ch{Nb} layer is sputtered at a rate of \SI{28}{\nm\per\min}, followed by an 1165 liftoff process. The wafer is then protected with resist and diced into \SI{10}{\mm}$\times$\SI{10}{\mm} dies for device processing.

The phononic crystal and release holes are then defined. Given the significant membrane size, proximity effect correction (PEC) was set up through BEAMER to address dose distortion. The pattern is subsequently exposed onto \SI{200}{\nm} CSAR resist in an EBL step. The resist is cold-developed in AR600-546, and the pattern is transferred through two consecutive dry etching steps: a \SI{50}{\nano\meter} aluminum etch using a \ch{Cl2/BCl3} chemistry, followed by a \SI{220}{\nano\meter} silicon etch using \ch{Cl2/HBr/O2} chemistry. The addition of \ch{O2} helps preserve the aluminum thin bridges and corners from being thinned and rounded during the silicon etch. The sample is immediately immersed in water to passivate the chlorinated aluminum, followed by resist stripping in \SI{80}{\degreeCelsius} 1165 remover for \SI{30}{\minute}.

Next, the microwave circuit is defined by patterning \SI{400}{\nm} PMMA A6 resist in an EBL step. To mitigate stitching errors, a \SI{10}{\micro \meter} field overlap is employed, along with a 2-multipass exposure configured using BEAMER. The resist is developed in MBIK/IPA at a 1:3 ratio, and the pattern is transferred by dry etching \SI{50}{\nano\meter} of Al and \SI{30}{\nano\meter} of Si. The silicon over-etching improves the surface for the Josephson junction evaporation. The sample is once again treated with water to passivate the chlorinated aluminum, and the resist is stripped by a \SI{30}{\minute} soak in \SI{80}{\degreeCelsius} 1165 remover.
\begin{figure*}[!htbp]
	\centering
	\includegraphics[width=2\columnwidth]{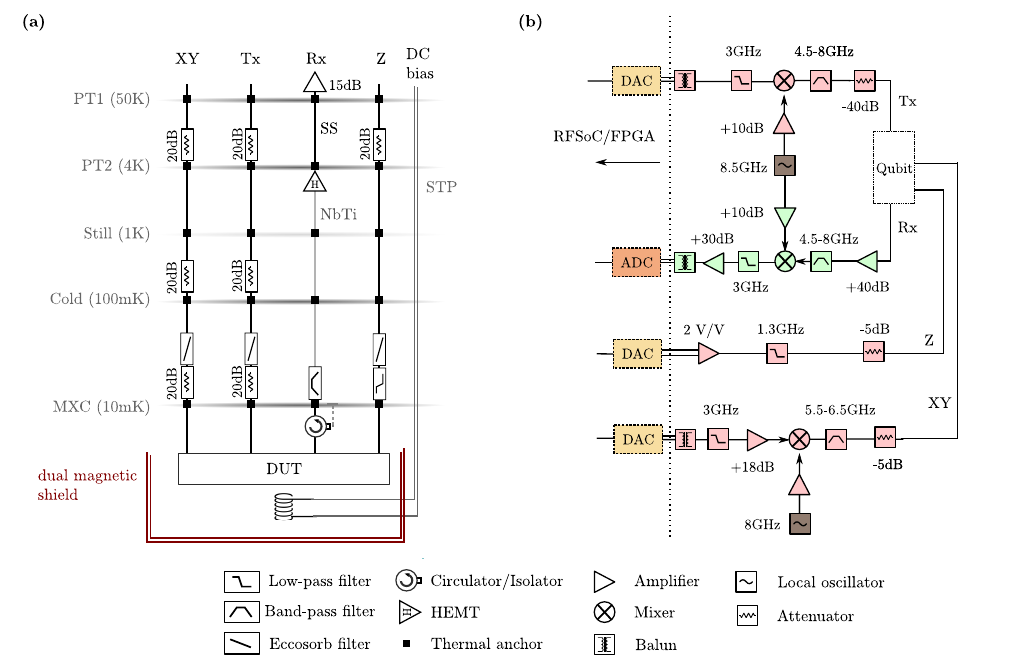}
	\caption{Experimental setup schematics for (a) dilution refrigerator cryogenic wiring, and (b) room temperature microwave electronics. }
	\label{SSetup}
\end{figure*}
The Josephson junctions are defined through EBL exposure of a \SI{400}{\nm}/\SI{200}{\nm} EL9/CSAR bilayer resist. The exposed resist is then sequentially cold-developed (MBIK-IPA 1:3/AR600-546) and gently descummed in \ch{O2} plasma. The sample is loaded into a double-angle evaporator (Plassys MEB550) and pumped down to a base pressure of \SI{4e-8}{\milli Torr} with the assistance of \ch{Ti} guttering. The subsequent steps are carried out in the following order: a \SI{30}{\nm} Al evaporation at coordinates ($\theta=45$, $\phi=45$); dynamic oxidation at \SI{20}{\milli \bar} for \SI{20}{\min}; another \SI{30}{\nm} Al evaporation at ($\theta=45$, $\phi=-90$); and a \SI{40}{\nm} Al evaporation at ($\theta=45$, $\phi=90$). The evaporation during all these steps is conducted at a rate of \SI{0.3}{\nm \per \sec}. The liftoff process is carried out by soaking the sample for \SI{2}{\hour} in a \SI{50}{\degreeCelsius} acetone bath, followed by a \SI{30}{\minute} soak in \SI{80}{\degreeCelsius} 1165 remover.

A second EBL step was employed to define a bandage layer. An ion milling process was used to remove the native oxide layer from the Al, followed by an Al evaporation step at ($\theta=0, \phi=0$), conducted at a rate of \SI{1}{\nm \per \sec}, resulting in a thickness of \SI{200}{\nm}. A liftoff and cleaning process similar to the one used in the JJ step was carried out. This step also served to increase the Z-line CPW thickness from \SI{50}{\nano \meter} to \SI{250}{\nano \meter}, allowing for a larger current capacity and avoiding heating issues. Aluminum wire bonds were used as airbridges to mitigate slot-line modes, which happen to be at lower frequencies than the main mode in released SOI CPW resonators. This step precedes the releasing process as wire bonding near suspended devices may induce structural collapse. The device is released using vapor HF through a \SI{4}{\um} isotropic oxide etch, conducted at a rate of \SI{36}{\nm \per \min}. Finally, the sample is mounted and wire-bonded onto a PCB enclosed by a copper box for measurement.

%####################
%####################
%####################
%####################
%####################

\subsection{Cryogenic setup}
We characterize the qubit in a \ch{^3He-^4He} dry dilution refrigerator (Bluefors, BF-LD250). This refrigerator comprises multiple temperature stages, namely PT1, PT2, Still flange, cold plate (CP), and mixing chamber (MXC) flange, as illustrated in Fig.~\ref{SSetup}(a). The Tx and Rx lines are used to probe the readout resonator, the XY line for qubit control, and the Z/DC-bias lines for dynamic/static control of the qubit's frequency. The Tx, XY, and Z lines pass through a series of cryogenic attenuators with a total of \SI{60}{\decibel}/\SI{60}{\decibel}/\SI{20}{\decibel} attenuation, respectively \cite{RFWiring}. All the input lines are filtered with Eccosorb IR filters (QMC-CRYOIRF-002MF-S), and an additional low-pass filter is added to the Z line (Minicircuits VLFX-1300+). The return Rx line passes through \SI{44}{\decibel} of isolation (2 x LNF-CIC4 8A), a bandpass filter (Keenlion KBF-4/8-2S), and is connected to a \SI{42}{\decibel} HEMT amplifier (LNF-LNC4 8C) through NbTi superconducting RF cable. The sample is mounted vertically inside a dual-cylinder magnetic shield (Cryo-Netic) along with an aligned superconducting coil made from NbTi DC wire for static biasing.

\subsection{Microwave electronics}
We use a Zynq UltraScale+ RFSoC board with the QICK (Quantum Instrumentation Control Kit) FPGA overlay for coherent RF signaling and acquisition up to \SI{3}{\giga \hertz} frequency \cite{QICK}. Front-end heterodyne stages with external local oscillators (LMX2595) parked at \SI{8.5}{\giga\hertz}/\SI{8}{\giga\hertz} for the Tx/XY lines are used for signal up-conversion to the desired readout/qubit frequency. For the Z-line, we employ a DC to \SI{800}{\mega \hertz} Differential-to-Single-Ended Opamp with a \SI{5000}{\volt \per \micro \second} slew rate for fast qubit tuning (TI THS3217), followed by a \SI{1.3}{\giga \hertz} low-pass filter (MC VLFX-1300). A series of amplifiers, filters, and attenuators are employed across the entire chain to ensure proper frequency mixing, sideband suppression, and utilization of the full DAC/ADC range, as illustrated in Fig.~\ref{SSetup}(b).

%####################
%####################
%####################
%####################
%####################

\end{document}